\documentclass[twocolumn,aps,superscriptaddress]{revtex4-1}
\usepackage{amsmath}
\usepackage{bm}
\usepackage{graphicx}
\usepackage{mathrsfs}
\usepackage{multirow}
\usepackage[normalem]{ulem}
\usepackage{graphicx}
\usepackage{booktabs, ctable}
\usepackage{xcolor, color, framed}
\usepackage[colorlinks, linkcolor=red,anchorcolor=green,citecolor=blue]{hyperref}

\graphicspath{{./figures/}}

\begin{document}

\title{Learning Hadron Emitting Sources with Deep Neural Networks}

\author{Lingxiao Wang}
\affiliation{Interdisciplinary Theoretical and Mathematical Sciences Program (iTHEMS), RIKEN Wako, Saitama 351-0198, Japan}

 \author{Jiaxing Zhao}
 \email{jzhao@itp.uni-frankfurt.de}
 \affiliation{Helmholtz Research Academy Hesse for FAIR (HFHF), GSI Helmholtz Center for Heavy Ion Physics, Campus Frankfurt, 60438 Frankfurt, Germany}
 \affiliation{Institut f\"ur Theoretische Physik, Johann Wolfgang Goethe-Universit\"at,Max-von-Laue-Straße 1, D-60438 Frankfurt am Main, Germany}

\date{\today}

\begin{abstract}
The correlation function observed in high-energy collision experiments encodes critical information about the emitted source and hadronic interactions. While the proton-proton interaction potential is well constrained by nucleon-nucleon scattering data, these measurements offer a unique avenue to investigate the proton-emitting source, reflecting the dynamical properties of the collisions. In this Letter, we present an unbiased approach to reconstruct proton-emitting sources from experimental correlation functions. Within an automatic differentiation framework, we parameterize the source functions with deep neural networks, to compute correlation functions. This approach achieves a lower chi-squared value compared to conventional Gaussian source functions and captures the long-tail behavior, in qualitative agreement with simulation predictions.
\end{abstract}

\maketitle

\emph{Introduction.--} Nuclear forces, also referred to as strong forces, are the forces that act between two or more nucleons (nuclei), which bind the nucleons (nuclei) together. 
The nuclear force between two nucleons is generated by mediating the $\pi$ meson, as first proposed by H. Yukawa~\cite{Yukawa:1935xg}. Following the experimental discovery of the heavier mesons, namely the $\sigma$, $\rho(770)$, and $\omega(782)$, the Yukawa theory was extended to the one-boson-exchange (OBE) model~\cite{Bryan:1964zzb,PhysRev.170.907}. 
There are numerous OBE-based/extended phenomenological nucleon-nucleon (N-N) potentials exist, including the Paris potential~\cite{Lacombe:1980dr}, the \textit{Argonne}-18 potential (Av-18)~\cite{Wiringa:1994wb}, the Reid Soft-Core potential~\cite{Reid:1968sq}, the Nijmegen potentials~\cite{Stoks:1994wp,Nagels:2014qqa}, and so on. The parameters in these forms are determined by explaining the N-N elastic scattering data.
Moreover, the chiral effective field theory approach is derived from the principles of Quantum Chromodynamics (QCD) using chiral perturbation theory to investigate the nuclear force~\cite{Entem:2003ft,Machleidt:2011zz,Epelbaum:2005pn,Drischler:2017wtt,Hammer:2012id,Gazit:2008ma}, which systematically consider the symmetries of QCD and provide a framework for including multi-nucleon forces. Additionally, the nuclear forces can also been extracted from the first-principles approach, namely lattice QCD~\cite{Aoki:2011ep,HALQCD:2018gyl,Yamazaki:2015asa,Wagman:2017tmp,Aoki:2023qih}. 

So far, we already have a better understanding of the N-N interaction. However, our comprehension of other interactions, such as those involving mesons and hyperons, remains limited. The clarification of these nuclear forces facilitates not only an understanding of the formation and reaction of nuclei, but also an understanding of the behavior of QCD matter at the most fundamental level. Meanwhile, the nuclear force plays a critical role in the evolution of stellar and supernova explosions, the formation of heavy elements in the universe, and the characteristics of neutron stars~\cite{Baym:2017whm}. 

\emph{Femtoscopy.--}
One effective experimental method for investigating hadronic interactions is the femtoscopy, which is inspired by the Hanbury Brown and Twiss (HBT) correlation~\cite{HBTc,Pratt:1984su,Lisa:2005dd}.
Femtoscopy method was utilized to probe the space-time configuration of the system at freeze-out~\cite{Lisa:2005dd,NA49:2007fqa,Li:2008qm,Wiedemann:1996ig,Kisiel:2006is,Xu:2024dnd}. The femtoscopy technique relates the final two-body correlation to the two-body interaction~\cite{Lisa:2005dd}. In accordance with the formalism of femtoscopy, the correlation function observed in the experiment can be calculated theoretically by convoluting the source function $S({\bm r})$ with the two-body scattering wavefunction $\psi_k({\bm r})$ via, 
\begin{eqnarray}
\label{eq.correlation}
C(k)=\int S({\bm r})|\psi_k({\bm r})|^2d{\bm r},
\end{eqnarray}
where $k=|{\bf p}_1-{\bf p}_2|/2$ is the relative momentum in the center-of-mass frame of the pair, and ${\bf r}$ is the relative distance between the two particles. The two-body scattering wave function $\psi_k({\bf r})$ can be obtained by solving the Schr\"odinger equation. This has been done in two popular tools that are the Correlation Afterburner (CRAB)~\cite{crabcite} and the Correlation Analysis Tool using the Schr\"odinger equation (CATS)~\cite{Mihaylov:2018rva,Fabbietti:2020bfg}.

Benefit from advanced experimental conditions, many two-hadrons correlations have been observed in both proton-proton and heavy-ion collisions in the Relativistic Heavy Ion Collider (RHIC)~\cite{STAR:2005rpl,STAR:2018uho,STAR:2014dcy,STAR:2015kha} and the Large Hadron Collider (LHC)~\cite{ALICE:2018ysd,ALICE:2021cpv,ALICE:2019buq,ALICE:2019hdt,ALICE:2022uso,ALICE:2011kmy}. This provides an excellent opportunity to study hadronic interactions. 
In experiments, the Lednick\'y and Lyuboshits model~\cite{Lednicky:1981su,Ohnishi:2016elb,ALICE:2018ysd} has consistently been employed to describe correlation functions, that assume a Gaussian source function and the interaction is encoded in the scattering length, $a_0$, and the effective range, $r_{\rm eff}$, which are based on a short-range interaction approximation. Nevertheless, the hadron-emitting source is beyond the Gaussian in the real case~\cite{Mihaylov:2018rva,ALICE:2023sjd}. The unknown source function will have a significant impact on the precision of the interaction. The proton-proton interaction is adequately addressed, allowing for precise investigation of the hadron-emitting source. This can further serve as a benchmark input, for probing hadronic interactions between other hadrons. 

\begin{figure}[!htb]
    \centering
    \includegraphics[width=0.47\textwidth]{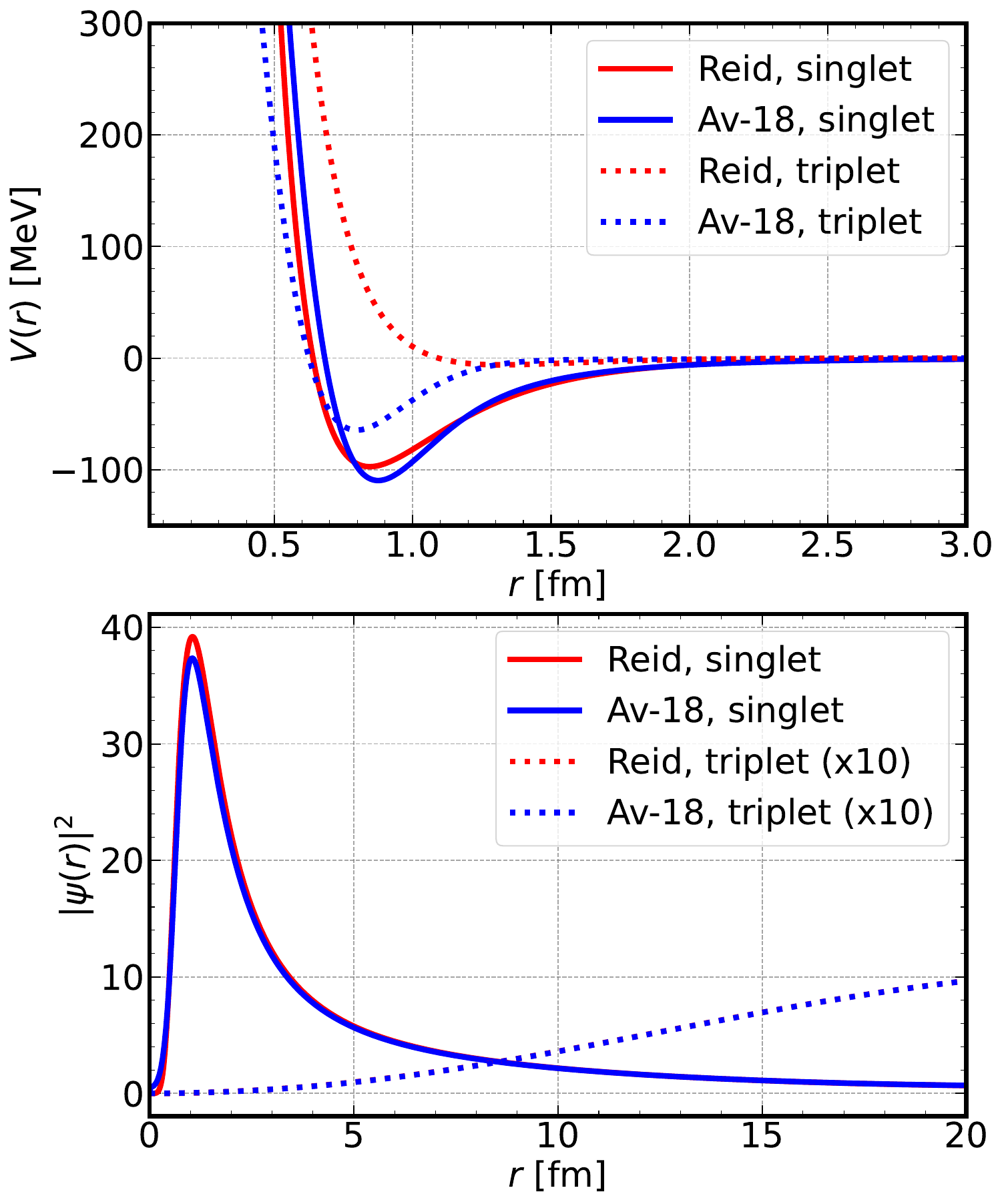}
    \caption{The proton-proton interaction potential (upper panel) and scattering wavefunction square (lower panel) at $k = 19.5$ MeV/c. The red and blue lines represent the Av-18 and Reid potential, respectively. The solid lines indicate the spin-singlet state, while the dashed lines depict the spin-triplet state.}
    \label{fig:wfpr}
\end{figure}

\emph{Interaction and scattering wavefunction.--}
As previously noted, the proton-proton interaction potential has been built by fitting the N-N scattering data. In this study, we consider two commonly used pp potentials: the Reid potential~\cite{Day:1981zz,Stoks:1994wp} and the Av-18 potential~\cite{Wiringa:1994wb}, as shown in the upper panel of Fig.~\ref{fig:wfpr}. These potentials include only strong interactions, with tensor interactions excluded. It can be observed that both the Av-18 and Reid potentials exhibit a pronounced repulsive core at short range, which gradually diminishes to zero at long range due to one-pion exchange. The intermediate range, defined as $1<r<2\rm fm$, is attributed to the exchange of scalar and vector mesons. Considering the spin configuration, the interaction potentials are categorized as either spin-singlet or spin-triplet. For the spin-singlet state, the Av-18 and Reid potentials are observed to be in close agreement. However, a significant difference is evident in the spin-triplet state, as depicted by the dotted lines.

In experiments, it is insufficient to distinguish between the spin-singlet and spin-triplet states, necessitating consideration of contributions from both. The relative contributions are determined by the spin degeneracy. Consequently, the correlation function given in Eq.~\eqref{eq.correlation} can be expressed as,
\begin{eqnarray}
\label{eq.correlationspin}
C(k) = \int S({\bm r}) \left( \frac{1}{4} |\psi_k^{S=0}({\bm r})|^2 + \frac{3}{4} |\psi_k^{S=1}({\bm r})|^2 \right) d{\bm r},
\end{eqnarray}
where the spin-singlet wave function $\psi_k^{S=0}({\bm r})$ and the spin-triplet scattering wave function $\psi_k^{S=1}({\bm r})$ are calculated by solving the Schr\"odinger equation with the aforementioned potential. The radial Schr\"odinger equation for the scattering states is,
\begin{eqnarray}
\frac{d^2 u_{k,l}^S(r)}{dr^2} = \left( 2\mu V(r) + \frac{l(l+1)}{r^2} - k^2 \right) u_{k,l}^S(r),
\end{eqnarray}
where $\mu = m_p / 2$ is the reduced mass, and $u_{k,l}^S(r)$ represents the radial wave function for spin state $S$ in  $l-$wave scattering. For low-energy scattering, the $S-$wave channel dominates. However, as the energy increases, contributions from higher-order spin channels become significant.

Consequently, the total scattering wavefunction can be expressed as follows,
\begin{eqnarray}
\label{eq.schroedingeq}
\psi_k^{S}({\bf r})=\sum_{l=0}^{l_{\rm max}}(2l+1)i^l{u_{k,l}^S(r)\over r}P_{l}(\cos \theta),
\end{eqnarray}
where $P_l$ denotes the Legendre polynomials. 
For low-energy scattering, the series converges relatively rapidly. In this study, we set $l_{\rm max} = 3$, and utilized the CATS to solve the Schr\"odinger equation~\cite{Mihaylov:2018rva,Fabbietti:2020bfg}. The convergence of the series was verified. The scattering wavefunctions are shown in the lower panel of Fig.~\ref{fig:wfpr}, where it is evident that the spin-singlet wavefunction is significantly larger than the spin-triplet wavefunction.

\begin{figure*}
    \centering
    \includegraphics[width=0.8\linewidth]{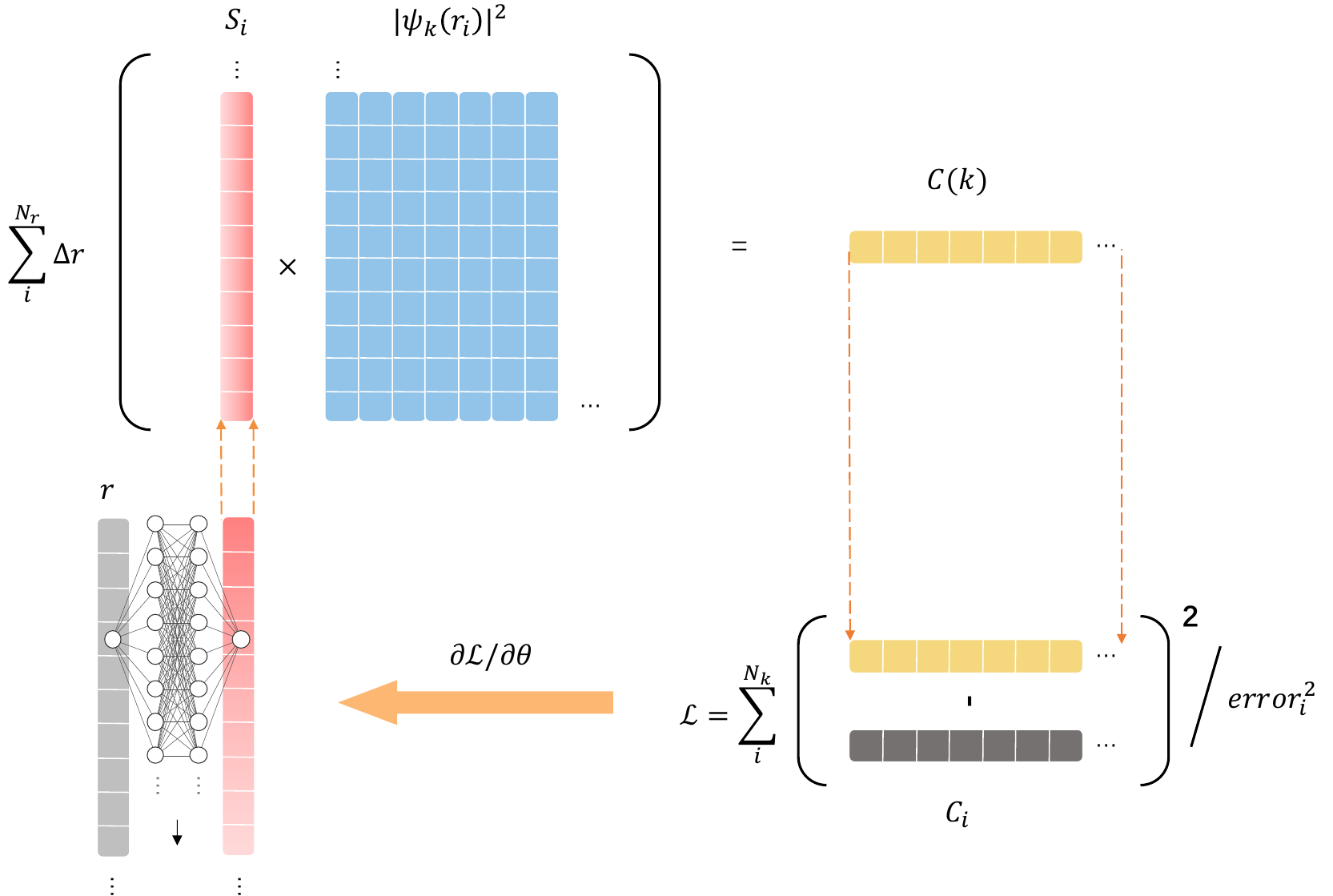}
    \caption{Build source functions from correlation functions with a neural network which has input and output nodes as $(r, S_{\boldsymbol{\theta}}(r))$ mapping.}
    \label{fig:flowchart_nn_source}
\end{figure*}

\emph{Building Source Functions.--}
The correlation function (CF), $C(k)$, can be obtained by convolving the scattering functions with the estimated source function (SF). The isotropic Gaussian function is the most commonly used form~\cite{Ohnishi:2023jlx}. Its functional form is given by Eq.~\eqref{eq.gaussian_source}, where the radius $r_0$ serves as the controlling parameter for the source size,
\begin{equation}
\label{eq.gaussian_source}
    S(r) = \frac{1}{(4\pi r_{0}^{2})^{3/2}}\exp\left(-\frac{r^{2}}{4r_{0}^{2}}\right).
\end{equation}
However, many studies have shown that the SF significantly deviates from the Gaussian \textit{Ansatz} with resonance contributions~\cite{Wiedemann:1996ig,Kisiel:2006is,ALICE:2023sjd}. This deviations is even more pronounced in small collision systems such as pp collisions~\cite{ALICE:2011kmy}. A solid modeling of the strong resonance contribution to the source function is imperative, yet currently lacking.
Such deviations can introduce non-negligible biases when attempting to understand hadron-hadron interactions through Femtoscopy. From an inverse problem perspective~\cite{Aarts:2025gyp}, we propose an automatic differentiable framework illustrated in Fig.~\ref{fig:flowchart_nn_source}, to reconstruct an unbiased source function directly from correlation functions. Here, an isotropic source approximation is adopted for $pp$ collisions, which is advantageous for smaller systems due to the absence of evident collective flow.

In our framework, the SF is from outputs of a deep neural network (DNN) \textit{Ansatz}, collected as $\vec{S} = [S_1, S_2, \cdots, S_{N_{r}}]$. It is illustrated in the lower-left panel of Fig.~\ref{fig:flowchart_nn_source}. The correlation function can be calculated as,
\begin{equation}
    C(k) = \sum_i^{N_{r}} S_i |\psi_k(r_i)|^2 \Delta r,
\end{equation}
where $N_{r}$ is the number of discrete radii and $\Delta r$ is the step size of the wavefunction. As shown in Fig.~\ref{fig:flowchart_nn_source}, after the forward process of the network and convolution, we obtain $\vec{S}$ and subsequently compute the reconstruction error as the loss function,
\begin{equation}
    \mathcal{L} = \sum_i^{N_{k}} {(\mathrm{C}_i - C(k_i))^2}/{\sigma_i^2},
    \label{eq.loss}
\end{equation}
where $\mathrm{C}_i$ represents the measured CF at $k_i$ with $N_k$ points, and $\sigma_i^2$ denotes the variance assigned to each uncorrelated observation in the standard $\chi^2$ function. Furthermore, this approach can be extended to multi-source observations by summing over them.

To optimize the parameters of the network representations, $\{\boldsymbol{\theta}\}$, with the loss function, we employ gradient-based algorithms. The gradient of the loss function is derived as,
\begin{align}
    \nabla_{\boldsymbol{\theta}} \mathcal{L} =
    \sum_{i,j}
    |\psi_{k_j}(r_i)|^2
    \frac{\partial \mathcal{L}}{\partial C(k_j)}
    \nabla_{\boldsymbol{\theta}} S_{\boldsymbol{\theta}}(r_i),
\end{align}
where $\nabla_{\boldsymbol{\theta}} S_{\boldsymbol{\theta}}(r_i)$ is computed using the standard backpropagation (BP) method in deep learning~\cite{bishop2023deep}. The reconstruction error is propagated through each layer of the neural network, and, combined with gradients derived via automatic differentiation, these are used to optimize the network parameters.

In details, an $L$-layer neural network is used to represent $S_{\boldsymbol{\theta}}(r)$, with an input node $r$ and a single output node $S_{\boldsymbol{\theta}}(r)$. The network comprises finite first-order differentiable modules, ensuring the continuity of the function $S_{\boldsymbol{\theta}}(r)$ is naturally preserved~\cite{2017arXiv170610239W,rosca:2020case,Shi:2022yqw}. We adopt a default parameter setting of $\text{width} = 64$ and $L = 4$ throughout this study. For optimizing the neural network representations, the Adam optimizer~\cite{kingma2014adam} is utilized, with a learning rate of $10^{-3}$ and training conducted over 10,000 epochs to approach the convergence. Additionally, a physical prior is embedded into the representations to ensure the positive definiteness of SFs~\cite{Wang:2021cqw,Shi:2021qri,Wang:2021jou,Aarts:2025gyp}. This is achieved by applying the \textit{Softplus} activation function at the output layer, defined as $\sigma(x) = \ln(1 + e^x)$.

\begin{figure}[!htb]
    \centering
    \includegraphics[width=0.45\textwidth]{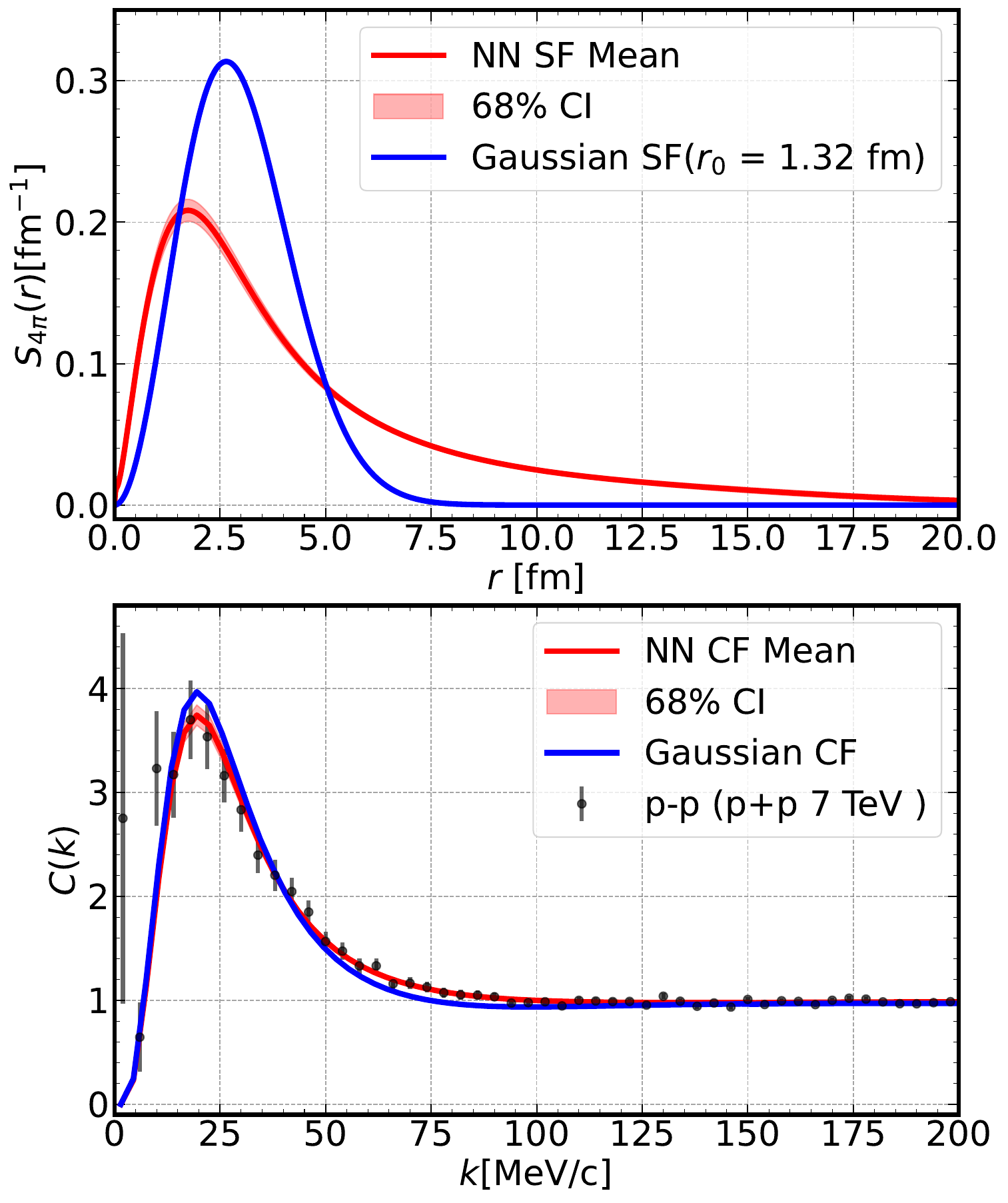}
    \caption{The correlation function of proton-proton pairs with a Gaussian source or Neural Network source, using the \textbf{Reid} potential, is shown in the bottom panel. The top panel presents the probability density function ($S_{4\pi}\equiv 4\pi r^2S(r)$) of the relative distance $r$ for hadron-hadron pairs produced. The red line corresponds to the Neural Network \textit{Ansatz}, while the blue line represents the Gaussian source with $r_0 = 1.32~\mathrm{fm}$. The experimental data are obtained from $pp$ collisions at $\sqrt{s} = 7~\mathrm{TeV}$ by the ALICE collaboration~\cite{ALICE:2018ysd}.}
    \label{fig:res_reid}
\end{figure}

\begin{figure}[!htb]
    \centering
    \includegraphics[width=0.45\textwidth]{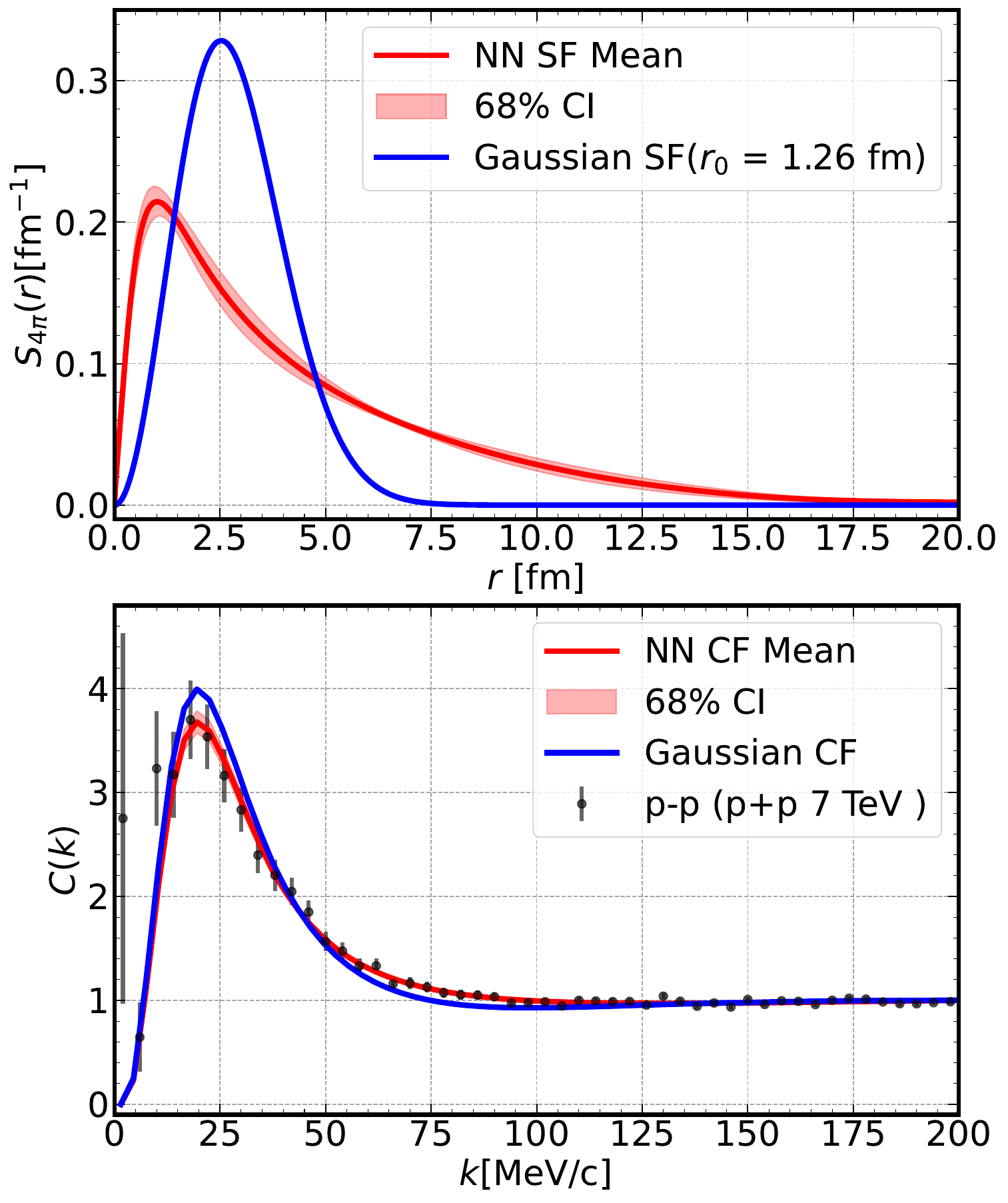}
    \caption{The correlation function of proton-proton pairs with a Gaussian source or Neural Network source, using the \textbf{Av-18} potential, is shown in the bottom panel. The top panel presents the probability density function of the relative distance $r$ for hadron-hadron pairs produced. The red line corresponds to the Neural Network \textit{Ansatz}, while the blue line represents the Gaussian source with $r_0 = 1.26~\mathrm{fm}$. The experimental data are obtained from $pp$ collisions at $\sqrt{s} = 7~\mathrm{TeV}$ by the ALICE collaboration~\cite{ALICE:2018ysd}.}
    \label{fig:res_av18}
\end{figure}

\emph{Correlation Functions.--}
The proton-proton CF can be obtained by convolving the scattering functions with the Gaussian source as Eq.~\eqref{eq.gaussian_source}, and the Neural Network source, as illustrated in Fig.~\ref{fig:res_reid} with the Reid potential and Fig.~\ref{fig:res_av18} with the Av-18 potential. 

The width $r_0$ of the Gaussian source is obtained by minimizing the chi-squared value relative to the experimental data, yielding $r_0 = 1.32~\mathrm{fm}$ for the Reid potential ($\chi^2 = 90.72$) and $r_0 = 1.26~\mathrm{fm}$ for the Av-18 potential ($\chi^2 = 89.12$). It is evident that the CF cannot be accurately reproduced across the full range of relative momentum in either case using the optimized Gaussian sources. In contrast, the SF constructed by the neural network ($\chi^2 = 34.55$ for the Reid potential and $\chi^2 = 35.81$ for the Av-18 potential) provides an adequate description of the experimental data, except for the first data point, which exhibits a significant degree of uncertainty and lies outside the predicted range. The uncertainty estimation is detailed in Appendix~\ref{ap.uncertainty}. A visual inspection of the SF reveals that it adopts a non-Gaussian form with a pronounced ``tail''. Such SF behavior has been observed in other simulations~\cite{Mihaylov:2018rva,ALICE:2023sjd,ALICE:2020ibs}. This marks the first time such behavior has been extracted directly from experimental measurements. These findings will provide critical input for theoretical simulations aimed at describing hadronic dynamics and serve as a benchmark for future hadronic femtoscopy studies.

\begin{figure}[!htb]
    \centering
    \includegraphics[width=0.45\textwidth]{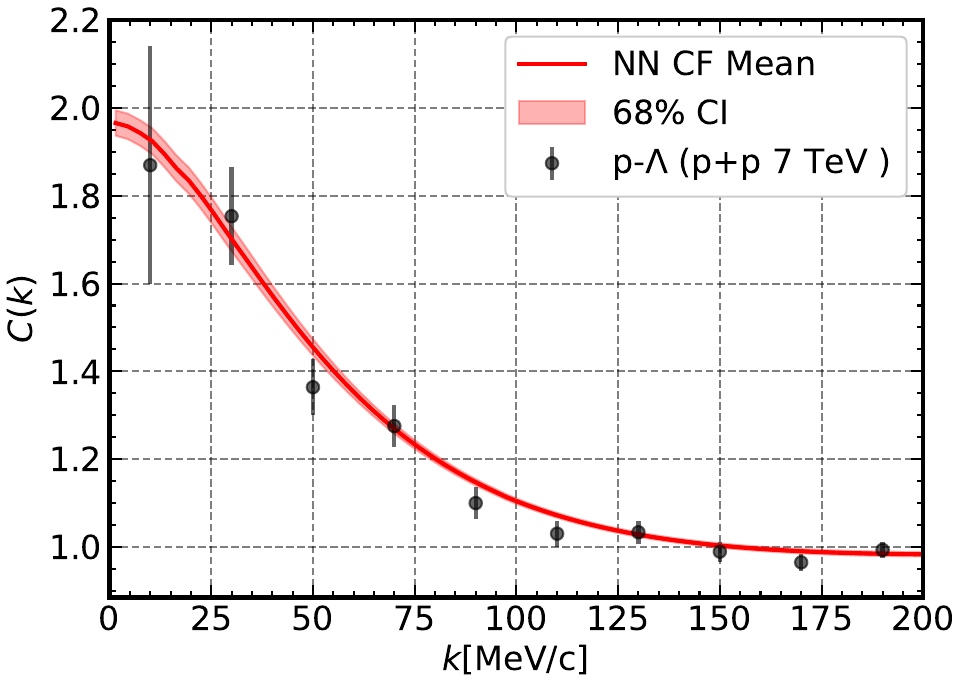}
    \caption{The correlation function of $p-\Lambda$ pairs is obtained using the Neural Network source, which was extracted based on the Av-18 potential derived from proton-proton correlations. The experimental data are from $pp$ collisions at $\sqrt{s} = 7~\mathrm{TeV}$, as measured by the ALICE collaboration~\cite{ALICE:2018ysd}.}
    \label{fig:corrplambda}
\end{figure}

The mass and quark composition of a hyperon is nearly identical to that of a proton. Accordingly, the learned hadron-emitting source can be utilized to study the hyperon–nucleon correlation and to test related hadronic interactions. The interaction potential between a proton and a hyperon $\Lambda$ has been described through many effective models~\cite{Bodmer:1984gc,Rijken:2006ep,Sasaki:2006cx,Polinder:2006zh,Nagels:2015lfa,Haidenbauer:2013oca,Nemura:2022wcs}. 
The results obtained from these models are rather different, but all confirm the attractiveness of the $p-\Lambda$ interaction for low momenta. With the learned hadron-emitting source, the correlation function of $p-\Lambda$ is computed assuming the interaction has a Usmani form~\cite{Bodmer:1984gc} with a $W_C=2250~\rm MeV$. This assumption gives a relative lower potential well that is similar to the preliminary lattice result~\cite{Nemura:2022wcs,Wang:1999bf}. A comparison with the experimental data is presented in Fig.~\ref{fig:corrplambda}. Therefore, we examine and impose constraints on the hyperon–nucleon interaction from an alternative perspective.

\emph{Summary.--} In this Letter, we extract the hadron-emitting source from the experimental proton-proton correlation function. The phenomenological proton-proton interaction is accurately described using a substantial body of nucleon-nucleon scattering data. The proton-proton scattering wavefunction is calculated with both the Reid and Av-18 potentials. The correlation functions are obtained by convolving the scattering wavefunction with the optimized Gaussian source function and the SF constructed using deep neural networks as an unbiased representation. Based on the experimental data, the latter produces a non-Gaussian SF, which exhibits a long-tail behavior. Notably, this work represents the first data-driven extraction of a hadron-emitting source, providing a more precise way to probe hadronic interactions in experiments through femtoscopy method. The hyperon–nucleon interaction is illustrated at the conclusion of this study. We will extend this study to extract the three-dimensional hadron-emitting source in heavy-ion collisions in future.   

\textbf{Acknowledgment.--} We thank Drs.\ Takumi Doi, Tetsuo Hatsuda, and Zhigang Xiao for helpful discussions.
We thank the DEEP-IN working group at RIKEN-iTHEMS for support in the preparation of this paper.
LW is supported by the RIKEN TRIP initiative (RIKEN Quantum).
JX is support by the Deutsche Forschungsgemeinschaft (DFG, German Research Foundation) through the grant CRC-TR 211 ’Strong-interaction matter under extreme conditions’ - Project number 315477589 - TRR 211.

\bibliographystyle{apsrev4-2}
\bibliography{refs}

\appendix

\section{Uncertainty Estimation}
\label{ap.uncertainty}
To evaluate the uncertainty of the reconstruction, a Bayesian perspective can be adopted, focusing on the posterior distribution of the SFs for the given astrophysical observations, $\text{Posterior}(\boldsymbol{\theta}_{\text{SF}}|\text{data})$. In this approach, an ensemble of $C(k_i)$ samples is first drawn from the normal distribution of real measurements. From this ensemble, the corresponding SFs are deterministically inferred using maximum likelihood estimation.  Given the ensemble of reconstructed SFs, importance sampling is applied to estimate the uncertainty associated with the desired posterior distribution. In this process, a proper weight is assigned to each SF to ensure accurate uncertainty quantification.

Our results and uncertainty estimations in the main text adhere to this strategy. In general, a physical variable $\hat{O}$ can be estimated as:
\begin{equation}
    \bar{O} = \langle \hat{O} \rangle = \sum_j^{N_\text{samples}} w^{(j)} O^{(j)}.
\end{equation}
The standard deviation is given by, $(\Delta O)^2 = \langle \hat{O}^2 \rangle - \bar{O}^2$. The weights are defined as similar in Ref.~\cite{Soma:2022vbb},
\begin{align}
    w^{(j)} &= \frac{\text{Posterior}(\boldsymbol{\theta}^{(j)}_{\text{SF}}|\text{data})}{\text{Proposal}(\boldsymbol{\theta}^{(j)}_{\text{SF}})} \\
    &\propto \frac{p(\text{data}|\boldsymbol{\theta}^{(j)}_{\text{SF}})\text{Prior}(\boldsymbol{\theta}^{(j)}_{\text{SF}})}{p(\boldsymbol{\theta}^{(j)}_{\text{SF}}|\text{samples}^{(j)})p(\text{samples}^{(j)}|\text{data})\text{Prior}(\text{data})}, \nonumber
\end{align}
where $j$ denotes the index of a reconstructed SF (from the sampled SF ensemble), and $\boldsymbol{\theta}_{\text{SF}}$ represents the parameter set describing the SF. 

Here, $p(\text{samples}|\text{data}) = \mathcal{N}(C_i, \Delta{C_i}^2)$ describes the probability of samples drawn from the normal distribution of errors. Additionally, $p(\boldsymbol{\theta}^{(j)}_{\text{SF}}|\text{samples}^{(j)}) = 1$, as the reconstruction deterministically locates the corresponding SF given the sampled $C(k_i)$ points. The likelihood function, $p(\text{data}|\boldsymbol{\theta}^{(j)}_{\text{SF}}) \propto \exp{(-\chi^2(C_{\boldsymbol{\theta}^{(j)}_{\text{SF}}})})$, quantifies the distances of the predicted $C(k_i)$ values from the real observations. For practical calculations, weights should be normalized as,
\begin{equation}
    \tilde{w}^{(j)} = \frac{w^{(j)}}{\sum_j w^{(j)}},
\end{equation}
and a cutoff is applied to mitigate the influence of outliers in the samples. In the normalization procedure, $\text{Prior}(\boldsymbol{\theta}^{(j)}_{\text{SF}}) $ and $\text{Prior}(\text{data})$ will be removed.

\end{document}